\documentclass[11pt, a4paper]{article}

\usepackage{latexsym}
\usepackage{graphicx}
\usepackage{cite}
\usepackage{amsmath}
\usepackage{amssymb}
\begin{document}
\title{Measurement of Static and Dynamic Light Scattering}
\author{Yong Sun{\footnote{Email: ysun200611@yahoo.ca}}\\
\emph{ysun200611@yahoo.ca}
\\\emph{Burnaby, BC, Canada}}
\maketitle

\begin{abstract}

Static Light Scattering (SLS) and Dynamic Light Scattering (DLS)
are the very important techniques to study the characteristics of 
materials in dispersion. One fundamental application is the accurate 
measurement of the size distribution. SLS measures using the the optical characteristic 
and DLS is determined by the hydrodynamic and optical
characteristics of particles in dispersion. In the theoretical
analysis of DLS, the relationship between SLS and DLS is given and
there exists an assumption that the static and hydrodynamic
radii are equal for spherical particles and then develop some theorems
to determine the structures of particles. Here, using the SLS technique, the size distribution 
can be measured accurately when the Rayleigh-Gans-Debye approximation is valid
for dilute homogenous spherical particles in dispersion. For the 
commercial samples, the static sizes are consistent with the sizes provided by
supplier respectively. The values of root mean square radius of gyration measured using the
Zimm plot and calculated using the commercial size distributions or size distributions measured using the SLS technique
are consistent respectively. Based on the static size distribution, with one assumption
between the static and hydrodynamic radii, the calculated and measured data of DLS 
are consistent very well at all the scattering angles investigated respectively.
Using the static size information the dimensionless shape parameter is discussed, our results show that
the shapes can not be determined based on the dimensionless shape parameter. Traditionally, the particle information:
apparent hydrodynamic radius and polydispersity index are obtained from DLS by analyzing the deviations of 
the intensity-intensity autocorrelation function from an exponent function. Since the apparent hydrodynamic 
radius obtained using the Stokes-Einstein relation is an optical weighted average radius, it is an approximate value 
of mean hydrodynamic radius. And the polydispersity index is not relate to the width of hydrodynamic radius. Therefore
the apparent hydrodynamic radius and polydispersity index cannot give an accurate description for size distribution.

\end{abstract}

\section{INTRODUCTION}

Light Scattering has been considered to be a well established
technique and applied in physics, chemistry, biology, etc. as an
essential tool to investigate the characteristics of nanoparticles
in dispersion. The treatment of Static Light Scattering (SLS)
spectroscopy is simplified to the Zimm plot, Berry plot or Guinier
plot etc. to get the root mean-square radius of gyration
$\left\langle R_{g}^{2}\right\rangle ^{1/2}$ and the molar mass of
particles provided that the particle sizes are small\cite{re1,re2}
and one fundamental application of Dynamic Light Scattering (DLS)
technique is the accurate and fast size measurement of particles
made of different materials. In general, the apparent
hydrodynamic radius and polydispersity index are obtained
using the method of Laplace transformation or
Cumulant analysis at a given scattering angle\cite{re3, re4, re5,
re6, re7}. Some researchers use the results measured using SLS and
DLS together to obtain much more information of nanoparticles
\cite{re8, re9, re10, re11, re12}. Although people have proposed
the theoretical function of the normalized time electric field
auto-correlation function of the scattered light $
g^{\left(1\right)}(\tau )$\cite{re4, re6, re13} and try to obtain
more information about nanoparticles using SLS and DLS together,
the comparison of the expected values and experimental data of the
normalized time auto-correlation function of the scattered light
intensity $g^{\left( 2\right) }\left( \tau \right) $ has not been
detailed.

How the particle size distribution can be obtained directly from
SLS data has been reported. Strawbridge and Hallett\cite{re14}
studied the theoretical scattered intensity of coated spheres with
vertically polarized incident light.  The scattered intensity at
the geometrical or linear trial radii between $r_{\min }$ and
$r_{\max }$ was used to fit SLS data. Schnablegger and
Glatter\cite{re15} assumed that the size distribution can be
described as a series of cubic B-splines and used the simulated
and measured data to demonstrate the computation procedure.

For dilute homogeneous spherical particles in dispersion, one method\cite{re19}
are used to obtain the accurate static size distribution. The
number distribution of particles is chosen as a Gaussian
distribution and the effects of the scattering vector and the
different intensity weights of different particle sizes on the
scattered light intensity are considered. With the assistance of a
non-linear least square fitting program, the mean static radius
$\left\langle R_{s}\right\rangle $ and standard deviation $\sigma
$ are measured accurately. For the polystyrene spherical particles investigated,
the values of mean static radius and mean radius measured using
Transmission Electron Microscopy (TEM) technique provided by the
supplier are consistent.

Based on the particle size distribution measured using SLS
technique, DLS technique was investigated further. The expected
values of $g^{\left( 2\right)}\left( \tau \right)$ calculated
based on the SLS measurements or commercial results provided by
the supplier are consistent with the experimental data very well.
This result also reveals that the static and hydrodynamic radii of
spherical nanoparticles are different and have a large difference. 

In order to obtain more information about particles, some people believe that
there exist some relationships between the physical quantities obtained using
SLS and DLS techniques. A lot of people believe that the dimensionless shape
parameter  ${\left\langle{R_g}^2\right\rangle^{\frac{1}{2}}}/{R_{app,h}}$ can give a good description of the 
shapes of particles. However based on our results, the dimensionless shape parameter
is ${\left\langle{R_g}^2\right\rangle^{\frac{1}{2}}}/{\left\langle{R_{s}}\right\rangle
}$. Therefore without the knowledge about 
the relationship of mean static and apparent hydrodynamic radii, this dimensionless
shape parameter ${\left\langle{R_g}^2\right\rangle^{\frac{1}{2}}}/{R_{app,h}}$ cannot be determined.

\section{THEORY}

When the homogeneous spherical particles are considered and
Rayleigh-Gans-Debye(RGD) approximation is valid, the average
scattered light intensity of a dilute non-interacting system in
unit volume can be obtained for vertically polarized light
\begin{equation}
\frac{I_{s}}{I_{inc}}=\frac{4\pi ^{2}\sin ^{2}\theta
_{1}n_{s}^{2}\left( \frac{dn}{dc}\right) _{c=0}^{2}c}{\lambda
^{4}r^{2}}\frac{4\pi \rho }{3} \frac{\int_{0}^{\infty
}R_{s}^{6}P\left( q,R_{s}\right) G\left( R_{s}\right)
dR_{s}}{\int_{0}^{\infty }R_{s}^{3}G\left( R_{s}\right) dR_{s}},
\label{mainfit}
\end{equation}
where $\theta _{1}$ is the angle between the polarization of the
incident electric field and the propagation direction of the
scattered field, $c$ is the mass concentration of particles, $r$
is the distance between the scattering particle and the point of
intensity measurement, $\rho $ is the density of particles,
$I_{inc}$ is the incident light intensity, $ I_{s}$ is the
intensity of the scattered light that reaches the detector, $
R_{s}$ is the static radius of a particle, $\ q=\frac{4\pi
}{\lambda }n_{s}\sin \frac{\theta }{2}$ is the scattering vector,
$\lambda $ is the wavelength of incident light in vacuo, $n_{s}$\
is the solvent refractive index, $ \theta $ is the scattering
angle, $P\left( q,R_{s}\right) $ is the form factor of homogeneous
spherical particles

\begin{equation}
P\left( q,R_{s}\right) =\frac{9}{q^{6}R_{s}^{6}}\left( \sin \left(
qR_{s}\right) -qR_{s}\cos \left( qR_{s}\right) \right) ^{2}
\label{P(qr)}
\end{equation}
and $G\left( R_{s}\right) $ is the number distribution of
particles. In this work, the number distribution is chosen as a
Gaussian distribution

\begin{equation}
G\left( R_{s};\left\langle R_{s}\right\rangle ,\sigma \right)
=\frac{1}{ \sigma \sqrt{2\pi }}\exp \left( -\frac{1}{2}\left(
\frac{R_{s}-\left\langle R_{s}\right\rangle }{\sigma }\right)
^{2}\right) ,
\end{equation}
where $\left\langle R_{s}\right\rangle $ is the mean static radius
and $\sigma $ is the standard deviation.

\noindent If the reflected light is considered, the average
scattered light intensity in unit volume is written as
\begin{equation}
\frac{I_{s}}{I_{inc}}=a\frac{4\pi \rho }{3}\frac{\int_{0}^{\infty
}R_{s}^{6}P\left( q,R_{s}\right) G\left( R_{s}\right)
dR_{s}+b\int_{0}^{\infty }R_{s}^{6}P\left( q^{\prime
},R_{s}\right) G\left( R_{s}\right) dR_{s}}{\int_{0}^{\infty
}R_{s}^{3}G\left( R_{s}\right) dR_{s}} \label{mainre}
\end{equation}
where

\begin{equation}
a=\frac{4\pi ^{2}\sin ^{2}\theta _{1}n_{s}^{2}\left(
\frac{dn}{dc}\right) _{c=0}^{2}c}{\lambda ^{4}r^{2}}
\end{equation}
and
\begin{equation}
q^{\prime }=\frac{4\pi }{\lambda }n_{s}\sin \frac{\pi -\theta }{2}
\end{equation}
is the scattering vector of the reflected light. $b$ is a constant
determined by the shape of sample cell, the refractive indices of
solvent and the sample cell and the geometry of instruments.

For dilute homogeneous spherical particles, $
g^{\left(1\right)}(\tau )$ can be obtained

\begin{equation}
g^{\left( 1\right) }\left( \tau \right) =\frac{\int R_{s}^{6}
P\left( q,R_{s}\right)G\left( R_{s}\right) \exp \left( -q^{2}D\tau
\right) dR_{s}}{\int R_{s}^{6}P\left( q,R_{s}\right) G\left(
R_{s}\right) dR_{s}}, \label{Grhrs}
\end{equation}
where $D$ is the diffusion coefficient.

\noindent From the Stokes-Einstein relation

\begin{equation}
D=\frac{k_{B}T}{6\pi \eta _{0}R_{h}},
\end{equation}
where $\eta _{0}$, $k_{B}$ and $T$ are the viscosity of solvent,
Boltzmann's constant and absolute temperature respectively, the
hydrodynamic radius $R_{h}$ can be obtained.

\noindent In this work, the relationship between the static and
hydrodynamic radii is assumed to be
\begin{equation}
R_{h}=kR_{s},  \label{RsRh}
\end{equation}
where $k$ is a constant. From the Siegert relation between
$g^{\left( 2\right) }\left( \tau \right) $ and $g^{\left( 1\right)
}\left( \tau \right)$ \cite{re7}

\begin{equation}
g^{\left( 2\right) }\left( \tau \right) =1+\beta \left( g^{\left(
1\right) }\right) ^{2},  \label{G1G2}
\end{equation}
the function between SLS and DLS is built and the values of $
g^{\left( 2\right) }\left( \tau \right) $ can be expected based on
the particle size information measured using SLS technique.

Traditionally for a solution of noninteracting monodisperse particles 
$g^{\left( 1\right) }\left( \tau \right) $ has this form
\begin{equation}
g^{\left( 1\right) }\left( \tau \right) =exp{\left(- \Gamma \tau \right)}
\end{equation}
where $\Gamma = q^2D$ is the decay rate, $D$ is the macromolecular translational
diffusion coefficient of the particles.
For a polydisperse system, $g^{\left( 1\right) }\left( \tau \right) $ consists of a distribution of exponentials
\begin{equation}
g^{\left( 1\right) }\left( \tau \right) = \int{G{\left(\Gamma \right)}exp{\left(- \Gamma \tau \right)}d\Gamma}
\end{equation}
where $G{\left(\Gamma \right)}$ is the normalized distribution of the decay rates.
The size distribution can be obtained using the method of moment analysis. The mean decay rate $\overline \Gamma$
and the moments of the distribution $\mu_2$ are defined as 
\begin{equation}
\overline \Gamma = \int{\Gamma G{\left(\Gamma \right)}d\Gamma}
\end{equation}
\begin{equation}
\mu_2 =  \int{{\left(\Gamma -\overline \Gamma\right)}^2 G{\left(\Gamma \right)}d\Gamma}
\end{equation}
The apparent hydrodynamic radius $R_{app,h}$ is defined using the Stokes-Einstein relation\cite{re20}
\begin{equation}
D=\frac{k_{B}T}{6\pi \eta _{0}R_{app,h}}=\overline \Gamma /q^2 ,
\end{equation}
The polydispersity index is defined as 
\begin{equation}
PD.I = \frac{\mu_2}{{\overline \Gamma}^2}.
\end{equation}
A dimensionless shape parameter $\rho$ of particles is defined as
\begin{equation}
\rho = {\left\langle{R_g}^2\right\rangle^{\frac{1}{2}}}/{R_{app,h}}
\end{equation}
For a long time, the measurements of $\rho$ are used to infer particle shapes.

\section{EXPERIMENT}

The SLS and DLS data were measured using the instrument built by
ALV-Laser Vertriebsgesellschaft m.b.H (Langen, Germany). It
utilizes an ALV-5000 Multiple Tau Digital Correlator and a JDS
Uniphase 1145P He-Ne laser to provide a 23 mW vertically polarized
laser at wavelength of 632.8 nm.

In this work, two kinds of samples were used. One is
Poly($N$-isopropylacrylamide) (PNIPAM) submicron spheres.
$N$-isopropylacrylamide (NIPAM, monomer) from Acros Organics was
recrystallized from hexane/acetone solution. Potassium persulfate
(KPS, initiator) and $N,N^{\prime }$-methylenebi-sacrylamide (BIS,
cross-linker) from Aldrich were used as received. Fresh de-ionized
water from a Milli-Q Plus water purification system (Millipore,
Bedford, with a 0.2 $\mu m$ filter) was used throughout the
experiments. The synthesis of gel particles was described
elsewhere\cite{re16,re17} and the recipes of the batches used in
this work are listed in Table \ref{table1}. The three samples are
named according to the molar ratios $n_{B}/n_{N}$ of $ N,N^{\prime
}$-methyle-nebisacrylamide over $N$-isopropylacrylamide.

\begin{center}
\begin{tabular}{|c|c|c|c|c|c|}
\hline Sample & T ($^\mathrm o$C) & $t$ (hrs) &
$ W_{N}+W_{B}$ (g) & $KPS$ (mg) & $n_{B}/n_{N}$ \\
\hline PNIPAM-1 & $70\pm 1$ & $4.0$ & $1.00$ & $40$ & $1.0\%$ \\
\hline PNIPAM-2 & $70\pm 1$ & $4.0$ & $1.00$ & $40$ & $2.0\%$ \\
\hline PNIPAM-5 & $70\pm 1$ & $4.0$ & $1.00$ & $40$ & $5.0\%$ \\
\hline
\end{tabular}
 \makeatletter\def\@captype{table}\makeatother
\caption{Synthesis conditions for PNIPAM particles.}
\label{table1}
\end{center}

The three PNIPAM samples were centrifuged at 14,500 RPM followed
by decantation of the supernatants and re-dispersion in fresh
de-ionized water. The process was repeated four times to remove of
free ions and any possible linear chains. Then the samples were
diluted for light scattering to weight factors of $8.56\times
10^{-6}$, $ 9.99\times 10^{-6}$ and $8.38\times 10^{-6}$ for
PNIPAM-1, PNIPAM-2 and PNIPAM-5 respectively. 0.45 $\mu m$ filters
(Millipore, Bedford) were used to clarify the samples before light
scattering measurements. The other kind of samples is two standard
polystyrene latex samples from Interfacial Dynamics Corporation
(Portland, Oregon). One polystyrene sample is the sulfate
polystyrene latex with a normalized mean radius of 33.5 nm
(Latex-1) and the other is the surfactant-free sulfate polystyrene
latex of 55 nm (Latex-2), as shown in Table \ref{table2}. Latex-1
and Latex-2 were diluted for light scattering to weight factors of
$1.02\times 10^{-5}$ and $1.58\times 10^{-5}$ respectively.

\section{DATA ANALYSIS}

In this section, how to measure the particle size distribution
using SLS technique and the comparison of the expected values and
experimental data of $g^{\left( 2\right) }\left( \tau \right) $
are shown.

\subsection{Standard polystyrene latex samples}

The particle size information was provided by the supplier as
obtained using TEM technique. Because of the small particle sizes
and the large difference between the refractive indices of the
polystyrene latex (1.591 at wavelength 590 nm and 20 $^\mathrm
o$C) and the water (1.332), i.e., the ``phase shift'' $\frac{4\pi
}{\lambda }R|m-1|$ \cite{re6, re18} are 0.13 and 0.21 for Latex-1
and Latex-2 respectively, which do not exactly satisfy the rough
criterion for validity of the RGD approximation\cite{re6}, the
mono-disperse model $G\left( R_{s}\right) =\delta \left(
R_{s}-\left\langle R_{s}\right\rangle \right) $ was used to
measure the approximate values of $\left\langle R_{s}\right\rangle
$ for the two polystyrene latex samples, respectively. The values
of the mean radii and standard deviations of the two samples shown
in Table \ref{table2} were input into Eq. \ref{mainfit} to get
$I_{s}/I_{inc}$ respectively. In order to compare with the
experimental data, the calculated value was set to be equal to
that of the experimental data at $q=0.0189$ nm$^{-1}$ for Latex-1.
This results and fitting results for Latex-2 are shown in Figs.
\ref{figPolyfitcal}a and \ref{figPolyfitcal}b, respectively. The
results show that the value measured using SLS technique is
consistent with that measured using TEM technique.

\begin{table}[ht]
\begin{center}
\begin{tabular}{|c|c|c|}
\hline $\left\langle R\right\rangle_{comm}$ (nm) & $\sigma_{comm}$
(nm) & $\left\langle R_{s}\right\rangle$ (nm)\\
\hline 33.5(Latex-1) & 2.5 & 33.3$\pm $0.2 \\
\hline 55(Latex-2) & 2.5 & 56.77$\pm $0.04 \\
\hline
\end{tabular}
\caption[]{The average radius measured using TEM and SLS
techniques.}\label{table2}
\end{center}
\end{table}

\begin{center}
  $\begin{array}{c@{\hspace{0in}}c}
     \includegraphics[width=0.35\textwidth,angle=-90]{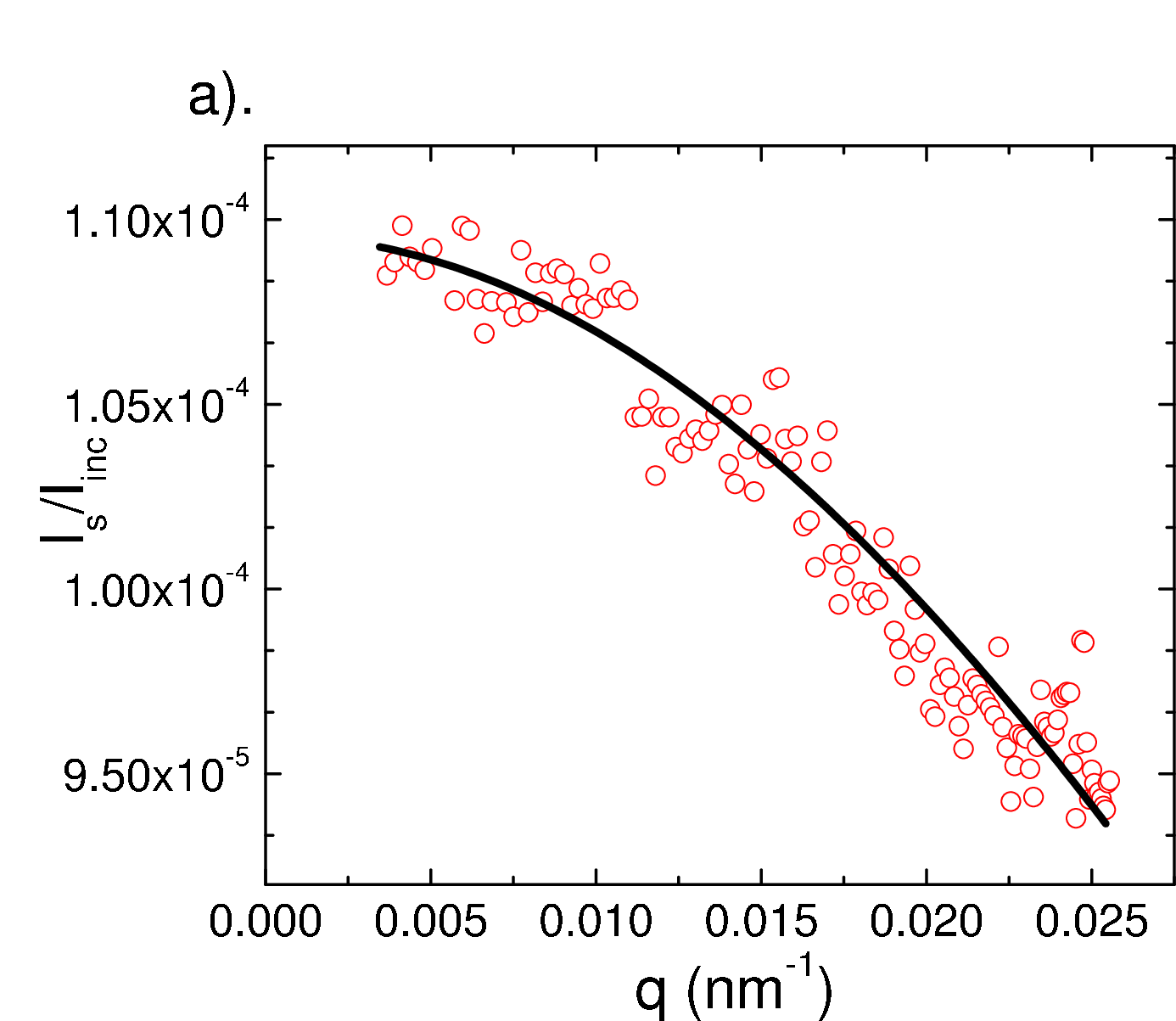} &
     \includegraphics[width=0.35\textwidth,angle=-90]{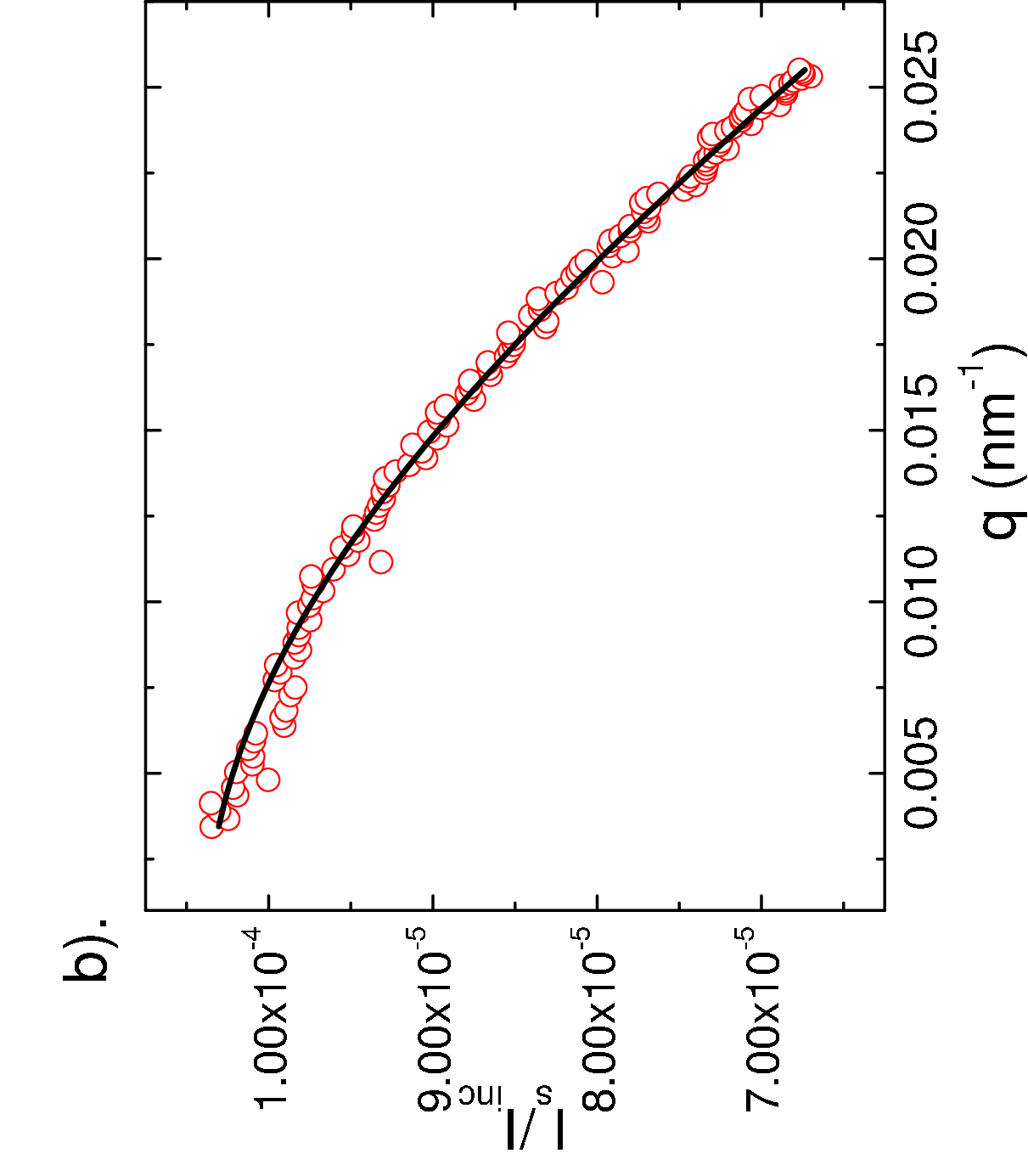} \\ [0.0cm]
    \end{array}$
   \end{center}\vspace{-0.5cm}
 \makeatletter\def\@captype{figure}\makeatother
\caption[] {a). The experimental data and expected values of
$I_{s}/I_{inc}$ for Latex-1 and b). The fitting results for
Latex-2. In a, the circles show the experimental data and the line
represents the expected values of $I_{s}/I_{inc}$. In b, the
circles show the experimental data and the line shows the fitting
results.} \label{figPolyfitcal}

If the constant $k$ in Eq. \ref{RsRh} is assumed to be 1.1 for
Latex-1 and 1.2 for Latex-2 and the size information provided by
the supplier is assumed to be consistent with that measured using
SLS technique, all the experimental data and expected values of
$g^{\left( 2\right) }\left( \tau \right) $ at the scattering
angles 30$^\mathrm o$, 60$^\mathrm o$, 90$^\mathrm o$,
120$^\mathrm o$ and 150$^\mathrm o$ and a temperature of 298.45 K
for Latex-1, 298.17 K for Latex-2 are shown in Figs.
\ref{figPolyDLScal}a and \ref{figPolyDLScal}b, respectively.
Figure \ref{figPolyDLScal} shows that the expected values are
consistent with the experimental data very well.

\begin{center}
  $\begin{array}{c@{\hspace{0in}}c}
     \includegraphics[width=0.35\textwidth,angle=-90]{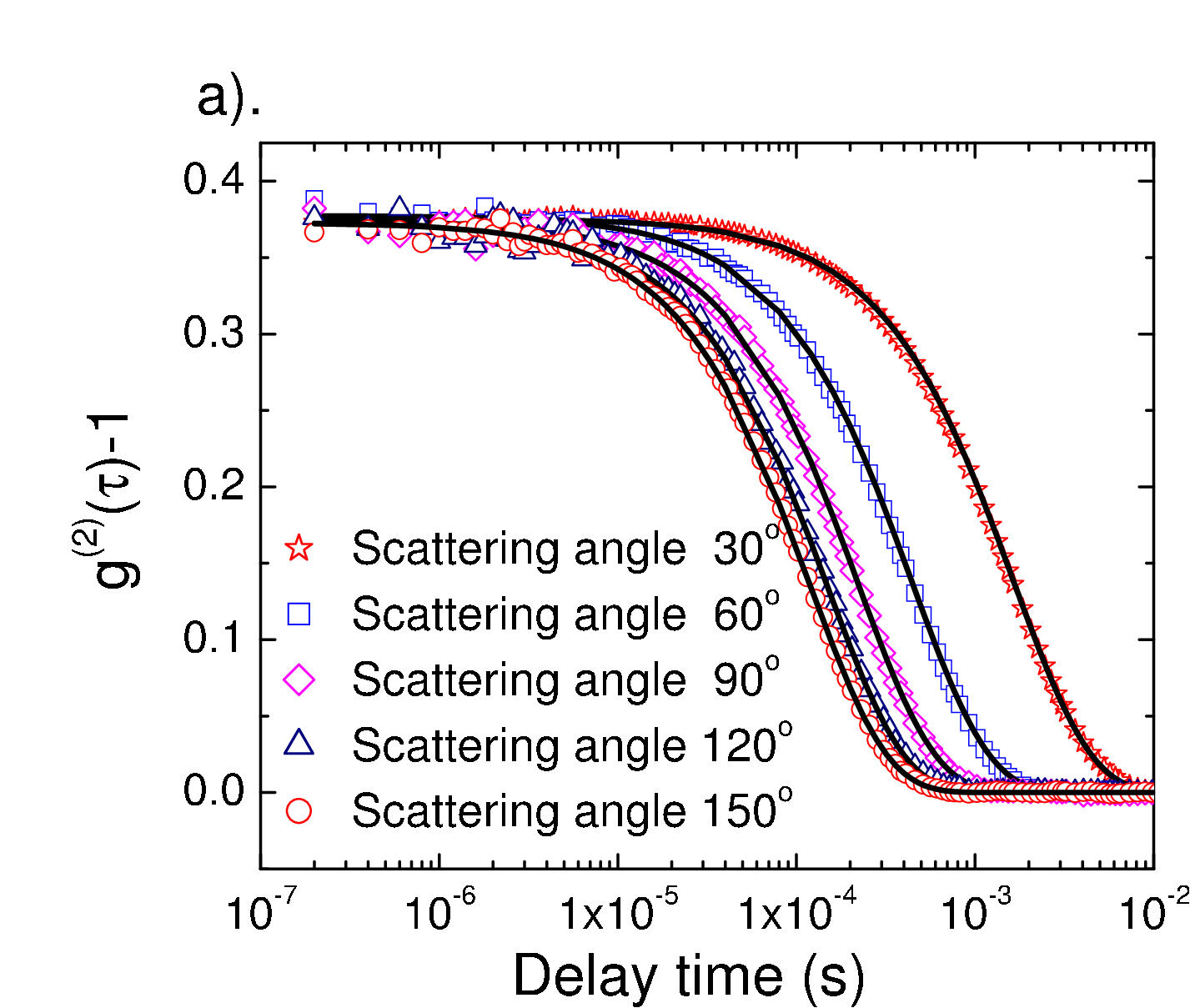} &
     \includegraphics[width=0.35\textwidth,angle=-90]{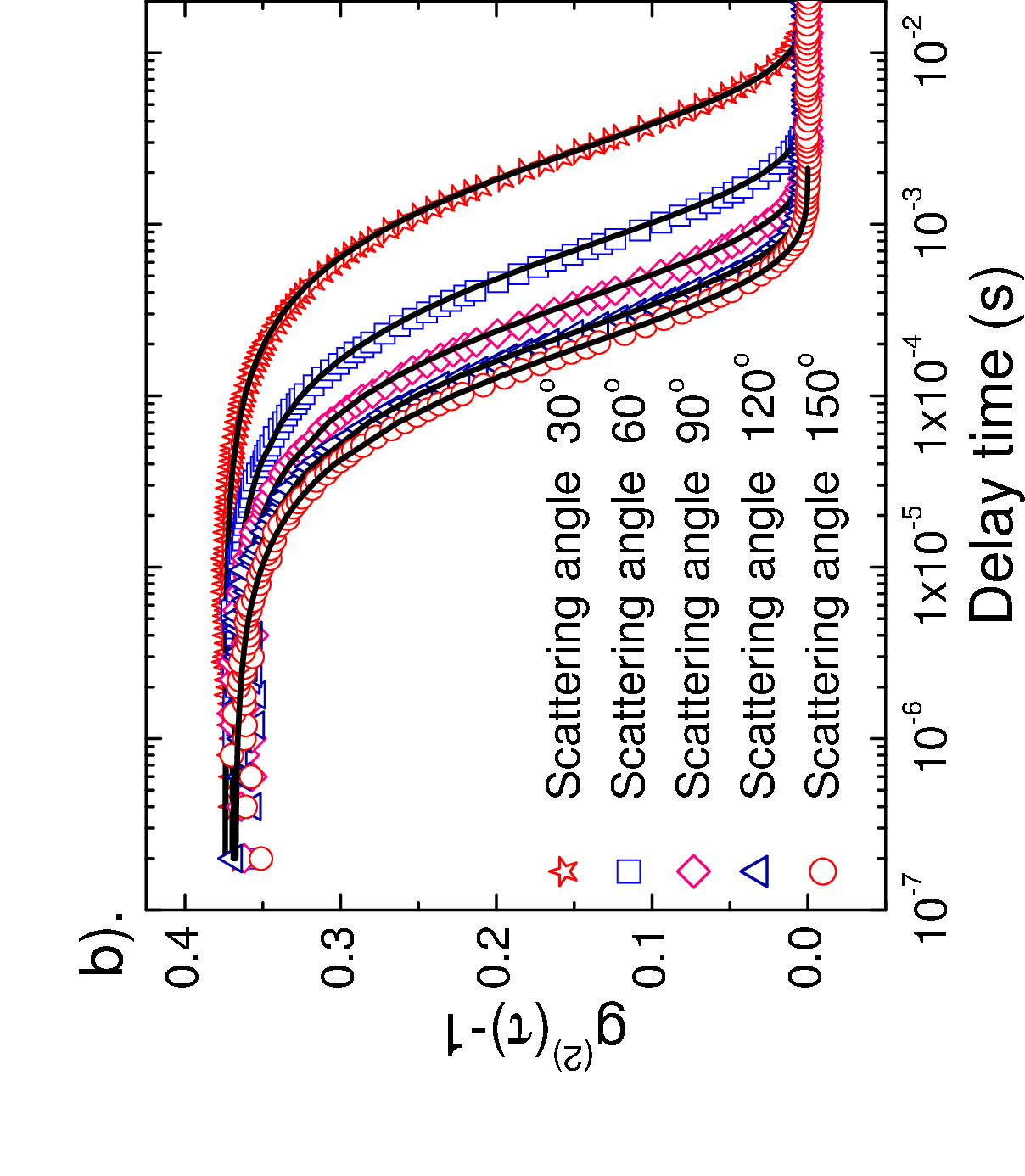} \\ [0.0cm]
    \end{array}$
   \end{center}\vspace{-0.5cm}
 \makeatletter\def\@captype{figure}\makeatother
  \caption{The experimental data and expected values of $g^{\left(
2\right) }\left( \tau \right) $. a). for Latex-1 and b). for
Latex-2. The symbols show the experimental data and the lines show
the expected values calculated under $R_{h}=kR_{s}$.}
\label{figPolyDLScal}

\subsection{PNIPAM samples}

When Eq. \ref{mainfit} was used to fit the SLS data of PNIPAM-1
measured at a temperature of 302.33 K, it was found that the
results of $\left\langle R_{s}\right\rangle $ and $\sigma $ depend
on the scattering vector range being fit, as shown in Table
\ref{table3}. If a small scattering vector range is chosen, the
parameters are not well-determined. As the scattering vector range
is increased, the uncertainties in the parameters decrease and
$\left\langle R_{s}\right\rangle $ and $\sigma $ stabilize. If the
fit range continues to increase, the values of $\left\langle
R_{s}\right\rangle $ and $\sigma $ begin to change and $\chi ^{2}$
grows. This is due to the deviation between the experimental and
theoretical scattered light intensity in the vicinity of the
scattered intensity minimum around $q=0.0177$ nm$^{-1}$, where
most of the scattered light is cancelled due to the light
interference. A lot of properties of nanoparticles could influence
the scattered light intensity in this region. For example, the
number distribution of particles deviates from a Gaussian
distribution, the particle shapes deviate from a perfect sphere
and the density of particles deviates from homogeneity, etc. In
order to avoid the influences of light interference, the stable
fit results $\left\langle R_s\right\rangle = 254.3 \pm 0.1$ nm and
$\sigma = 21.5 \pm 0.3$ nm obtained in the scattering vector range
from 0.00345 to 0.01517 nm$^{-1}$ are chosen as the size
information measured using SLS technique. The stable fit results
and the residuals in the scattering vector range from 0.00345 to
0.01517 nm$^{-1}$ are shown in Fig. \ref{figPNI-302fit}.

\begin{center}
\begin{tabular}{|c|c|c|c|}
\hline $q$ ($10^{-3}$ nm$^{-1}$) & $\left\langle
R_{s}\right\rangle$ (nm) & $ \sigma$ (nm) & $\chi ^{2}$ \\
\hline 3.45 to 9.05 & 260.09$\pm $9.81 & 12.66$\pm $19.81 & 1.64 \\
\hline 3.45 to 11.18 & 260.30$\pm $1.49 & 12.30$\pm $3.37 & 1.65 \\
\hline 3.45 to 13.23 & 253.45$\pm $0.69 & 22.80$\pm $0.94 & 2.26 \\
\hline 3.45 to 14.21 & 254.10$\pm $0.15 & 21.94$\pm $0.36 & 2.03 \\
\hline 3.45 to 15.17 & 254.34$\pm $0.12 & 21.47$\pm $0.33 & 2.15 \\
\hline 3.45 to 17.00 & 255.40$\pm $0.10 & 17.32$\pm $0.22 & 11.02 \\
\hline
\end{tabular}
 \makeatletter\def\@captype{table}\makeatother
\caption{The fit results obtained using Eq. \ref{mainfit} for
PNIPAM-1 at different scattering vector ranges and a temperature
of 302.33 K.}\label{table3}
\end{center}

\begin{center}
   \includegraphics[width=0.35\textwidth,angle=-90]{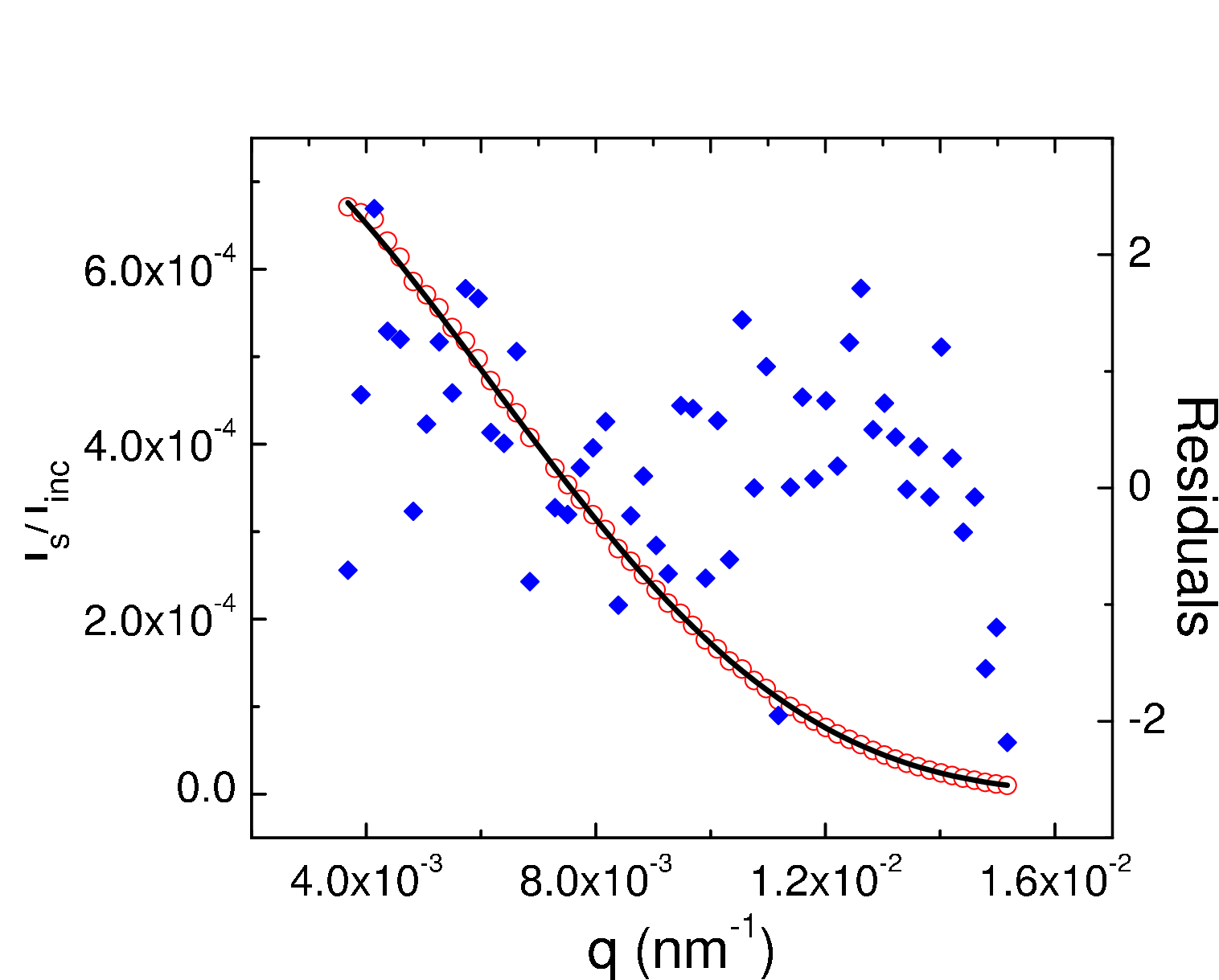}
      \makeatletter\def\@captype{figure}\makeatother
   \caption{The experimental data and stable fit results obtained using
   Eq. \ref{mainfit} for PNIPAM-1 at a temperature of 302.33 K.
   The circles show the experimental data, the line shows the fit results and the diamonds show the
residuals: $\left( y_{i}-y_{fit}\right) /\sigma _{i}$.}
 \label{figPNI-302fit}
\end{center}

When the reflected light was considered, Eq. \ref{mainre} was used
to fit the data in the full scattering vector range (0.00345 to
0.0255 nm$^{-1}$) for the various factors of reflected light $b$.
The fit results are listed in Table \ref{table4}. The results show
that the values of $\chi ^{2}$ are much larger.  The mean static
radius $\left\langle R_{s}\right\rangle $ is consistent with that
measured using Eq. \ref{mainfit} in the scattering vector range
from 0.00345 to 0.01517 nm$^{-1}$ and the standard deviation
changes to smaller.

\begin{center}
\begin{tabular}{|c|c|c|c|}
\hline $b$ & $\left\langle R_{s}\right\rangle$ (nm) & $\sigma$
(nm) & $\chi ^{2}$ \\
\hline 0.01 & 254.0$\pm $0.3 & 14.4$\pm $0.5 & 194.60 \\
\hline 0.011 & 254.0$\pm $0.3 & 14.6$\pm $0.5 & 168.20 \\
\hline 0.012 & 254.0$\pm $0.3 & 14.7$\pm $0.5 & 149.99 \\
\hline 0.013 & 254.0$\pm $0.2 & 14.8$\pm $0.4 & 139.82 \\
\hline 0.014 & 254.1$\pm $0.2 & 15.0$\pm $0.4 & 137.52 \\
\hline 0.015 & 254.1$\pm $0.2 & 15.1$\pm $0.4 & 142.96 \\
\hline 0.016 & 254.09$\pm $0.07 & 15.2$\pm $0.5 & 155.97 \\
\hline 0.017 & 254.1$\pm $0.3 & 15.4$\pm $0.5 & 176.40 \\
\hline 0.018 & 254.1$\pm $0.3 & 15.5$\pm $0.5 & 204.08 \\
\hline
\end{tabular}
 \makeatletter\def\@captype{table}\makeatother
\caption{The fit results for PNIPAM-1 obtained from Eq.
\ref{mainre} using the various values of $b$.}\label{table4}
\end{center}

As discussed above, light interference in the vicinity of the
scattered intensity minimum would influence the fit results. In
order to eliminate the effects of light interference, the
experimental data in the vicinity of the scattered intensity
minimum were neglected. Eq. \ref {mainre} was thus used to fit the
experimental data in the full scattering vector range again. The
fit values are listed in Table \ref{table5}. The mean static
radius and standard deviation are consistent with the stable fit
results obtained using Eq. \ref {mainfit} in the scattering vector
range from 0.00345 to 0.01517 nm$^{-1}$.

\begin{center}
\begin{tabular}{|c|c|c|c|}
\hline $b$ & $\left\langle R_{s}\right\rangle$ (nm) & $\sigma$
(nm) & $\chi ^{2}$ \\
\hline 0.013 & 251.3$\pm $0.6 & 22.17$\pm $0.05 & 79.80 \\
\hline 0.014 & 251.1$\pm $0.6 & 23.3$\pm $0.9 & 58.29 \\
\hline 0.015 & 250.9$\pm $0.6 & 24.4$\pm $0.8 & 44.50 \\
\hline 0.016 & 250.7$\pm $0.5 & 25.4$\pm $0.7 & 37.02 \\
\hline 0.017 & 250.5$\pm $0.6 & 26.4$\pm $0.7 & 36.01 \\
\hline 0.018 & 250.3$\pm $0.6 & 27.24$\pm $0.8 & 41.59 \\
\hline
\end{tabular}
 \makeatletter\def\@captype{table}\makeatother
\caption{The fit results for PNIPAM-1 obtained using Eq.
\ref{mainre} and neglecting experimental data near the intensity
minimum.}\label{table5}
\end{center}

If the constant $k$ in Eq. \ref{RsRh} for the PNIPAM-1 is assumed
to be 1.21, all the experimental data and expected values of
$g^{\left( 2\right) }\left( \tau \right) $ at the scattering
angles 30$^\mathrm o$, 50$^\mathrm o$ and 70$^\mathrm o$ are shown
in Fig. \ref{figPNI-302cal}. Figure \ref{figPNI-302cal} shows that
the expected values are consistent with the experimental data very
well.

\begin{center}
  \includegraphics[width=0.35\textwidth,angle=-90]{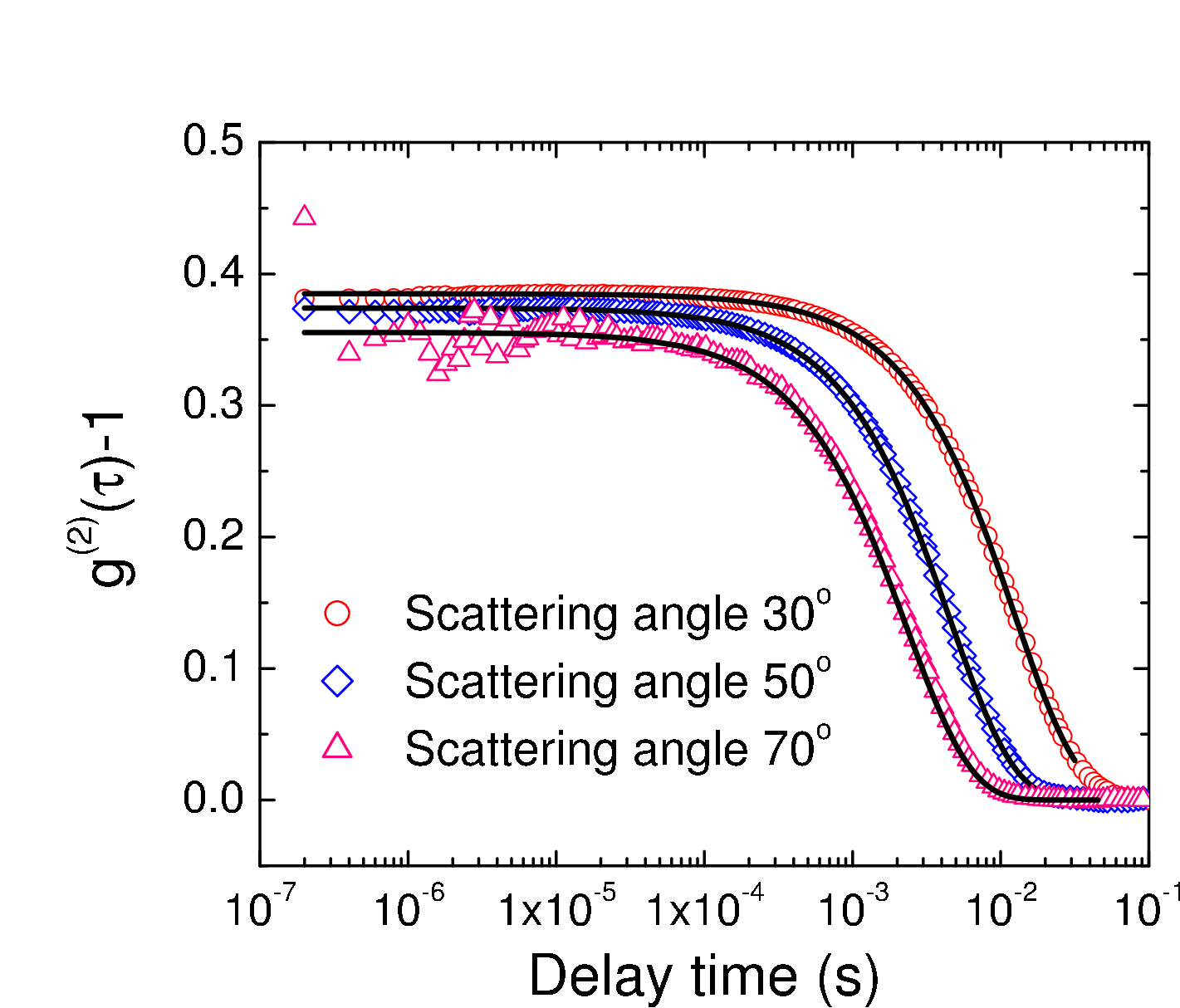}
   \makeatletter\def\@captype{figure}\makeatother
  \caption{The experimental data and expected values of $g^{\left(
2\right) }\left( \tau \right)$ for PNIPAM-1 at a temperature of
302.33 K. The symbols show the experimental data and the lines
show the expected values calculated under $R_{h}=1.21R_{s}$.}
\label{figPNI-302cal}
\end{center}

For the PNIPAM samples at high temperatures, the situation using
Eq. \ref{mainfit} is the same as that of PNIPAM-1 at a temperature
of 302.33 K. The values of $\left\langle R_{s}\right\rangle $ and
$\sigma $ depend on the scattering vector range being fit. If a
small scattering vector range is chosen, the parameters are not
well-determined. As the scattering vector range is increased, the
uncertainties in the parameters decrease and $\left\langle
R_{s}\right\rangle $ and $\sigma $ stabilize. The stable fit
results $\left\langle R_s\right\rangle = 139.3 \pm 0.3$ nm and
$\sigma = 12.4 \pm 0.6$ nm obtained in the scattering vector range
from 0.00345 to 0.02555 nm$^{-1}$ for PNIPAM-5 at a temperature of
312.66 K are chosen as the size information measured using SLS
technique. Figure \ref{figPNIhT}a shows the stable fit results and
the residuals. If the constant $k$ in Eq. \ref{RsRh} for PNIPAM-5
is assumed to be 1.1, all the experimental and expected values of
$g^{\left( 2\right) }\left( \tau \right) $ at the scattering
angles 30$^\mathrm o$, 50$^\mathrm o$, 70$^\mathrm o$ and
100$^\mathrm o$ are shown in Fig. \ref{figPNIhT}b. The expected
values of $g^{\left( 2\right) }\left( \tau \right) $ are
consistent with the experimental data very well.

\begin{center}
  $\begin{array}{c@{\hspace{0in}}c}
     \includegraphics[width=0.35\textwidth,angle=-90]{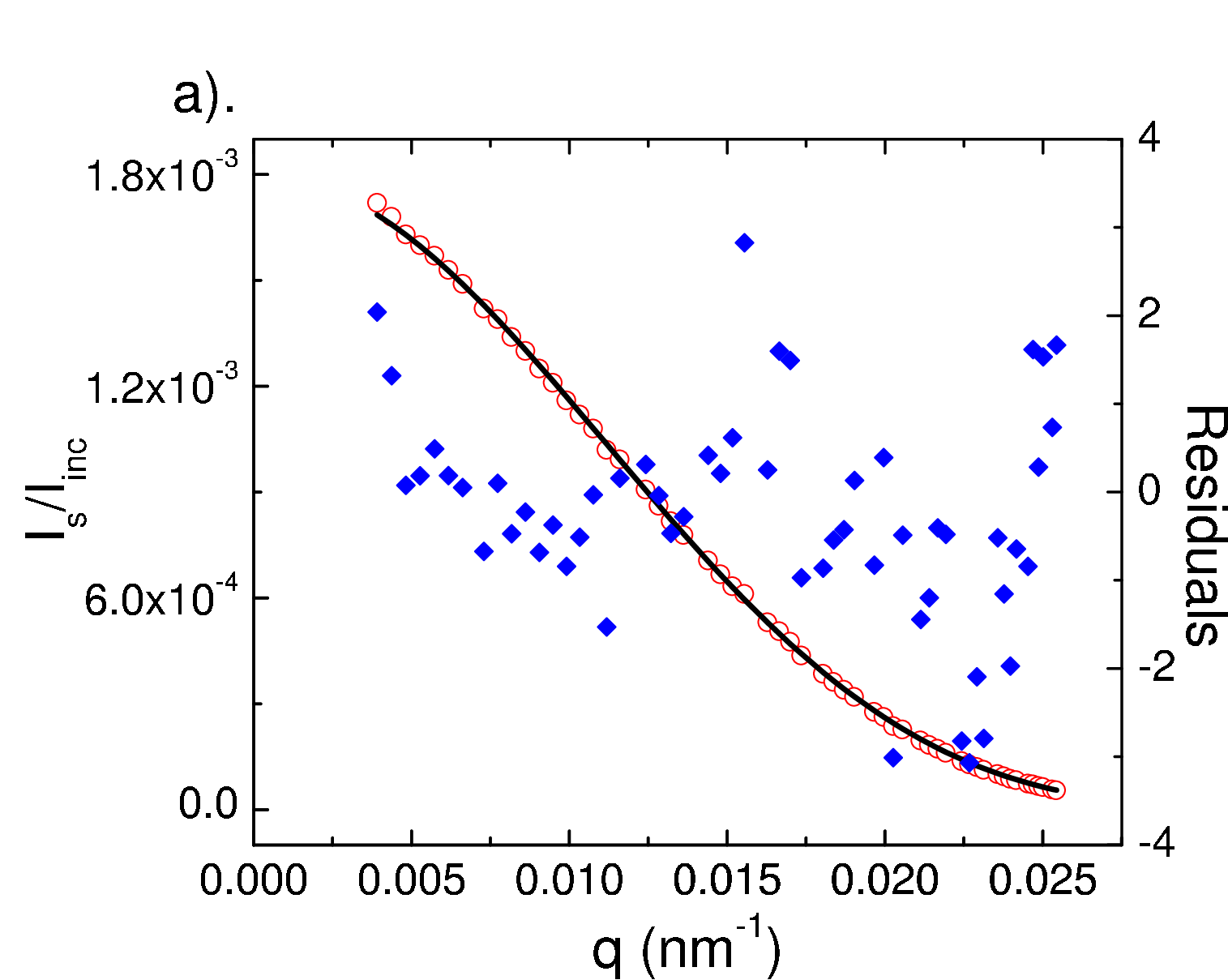} &
     \includegraphics[width=0.35\textwidth,angle=-90]{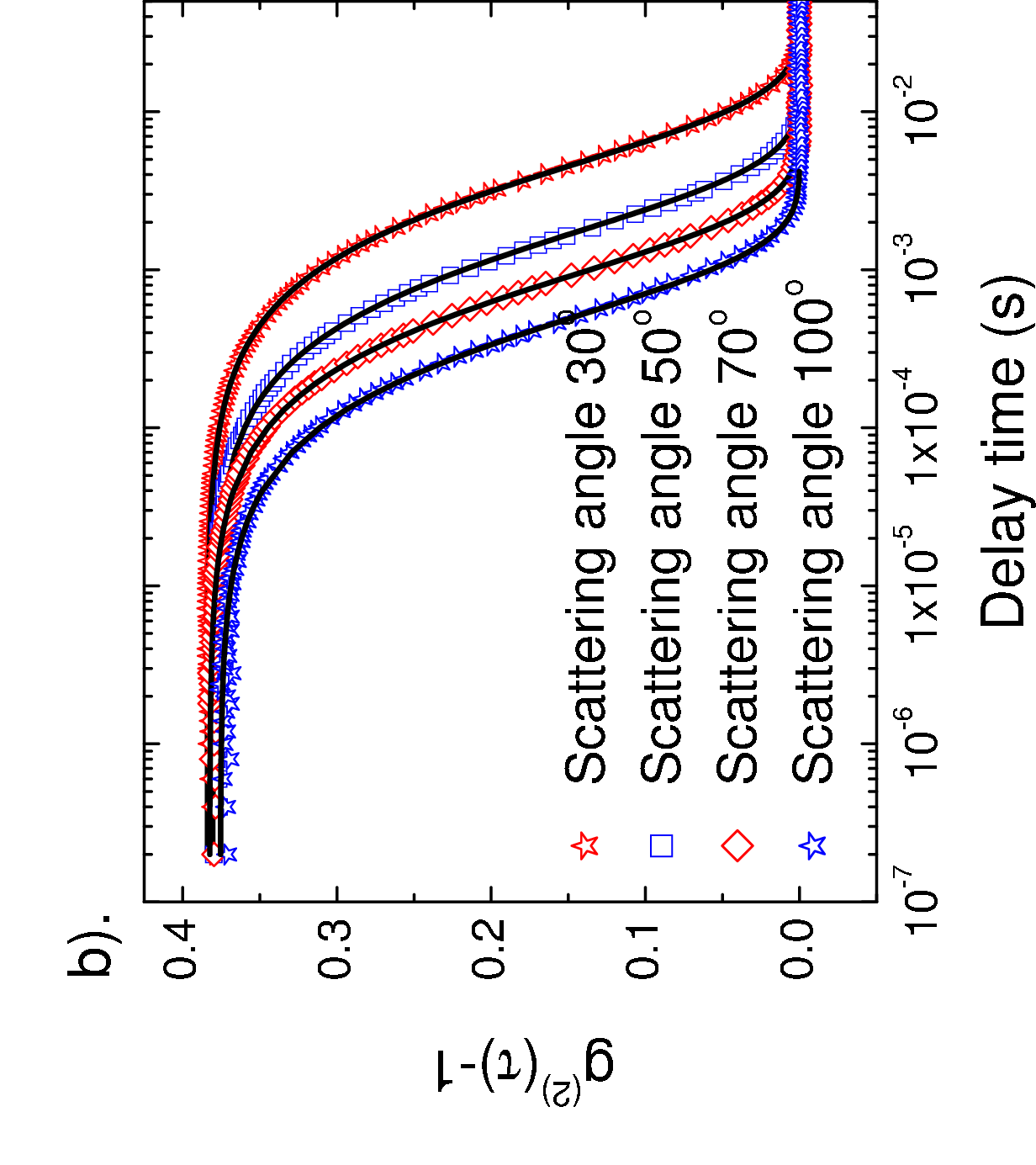} \\ [0.0cm]
    \end{array}$
   \end{center}\vspace{-0.5cm}
 \makeatletter\def\@captype{figure}\makeatother
  \caption{The results of PNIPAM-5 at a temperature 312.66 K. a). The experimental data and stable fit results obtained using
   Eq. \ref{mainfit}. The circles show the experimental data, the line shows the fit results and the diamonds show the
residuals: $\left( y_{i}-y_{fit}\right) /\sigma _{i}$. b). The
experimental data and expected values of $g^{\left( 2\right)
}\left( \tau \right)$. The symbols show the experimental data and
the lines show the expected values calculated under
$R_{h}=1.1R_{s}$.} \label{figPNIhT}

\section{RESULTS AND DISCUSSION}

Same conclusions are also obtained for all other samples
investigated. The fit results of $\left\langle R_{s}\right\rangle
$ and $\sigma $ depend on the scattering vector range being fit.
If a small scattering vector range is chosen, the parameters are
not well-determined. As the scattering vector range is increased,
the uncertainties in the parameters decrease and $\left\langle
R_{s}\right\rangle $ and $\sigma $ stabilize.

For the PNIPAM samples, the main reason for the difference between
experimental and theoretical scattered intensity in the vicinity
of the scattered intensity minimum seems to be that the small and
large particles in solution do not exist. With the mean static
radius and standard deviation obtained using Eq. \ref{mainfit} in
the $q$ range between 0.00345 and 0.01517 nm$^{-1}$, three
different ways of calculation were performed to explore which can
give the best expectation of the experimental data. In Fig.
\ref{figdiffcal}, the expected values of the scattered intensity
related to incident intensity were first calculated from Eq.
\ref{mainfit} in the full particle size distribution range between
1 and 800 nm. The calculated curve (solid line) matches the
experimental date points only when q is smaller than 0.016
nm$^{-1}$. Then, a truncated Gaussian distribution was used and
the calculation was performed between $\left\langle
R_{s}\right\rangle -1.3\sigma $ and $\left\langle
R_{s}\right\rangle +1.3\sigma $ using Eq. \ref{mainfit}. The
calculated curve (dash line) matches the experimental date points
in a broader $q$ range including the vicinity of the scattered
intensity minimum and deviates only at $q \geq 0.021$ nm$^{-1}$
where the reflected light could be detected. Finally, the
integrated range did not change but the reflected light was
considered and Eq. \ref{mainre} was used to calculate the expected
results assuming $b=0.014$. The calculated curve (dot line)
matches the experimental date in all $q$ range investigated. The
results show that the scattered intensity in the vicinity of the
scattered intensity minimum is very sensitive to the particle size
distribution and the influences of the reflected light only lies
at very large scattering vectors.

\begin{center}
   \includegraphics[width=0.35\textwidth,angle=-90]{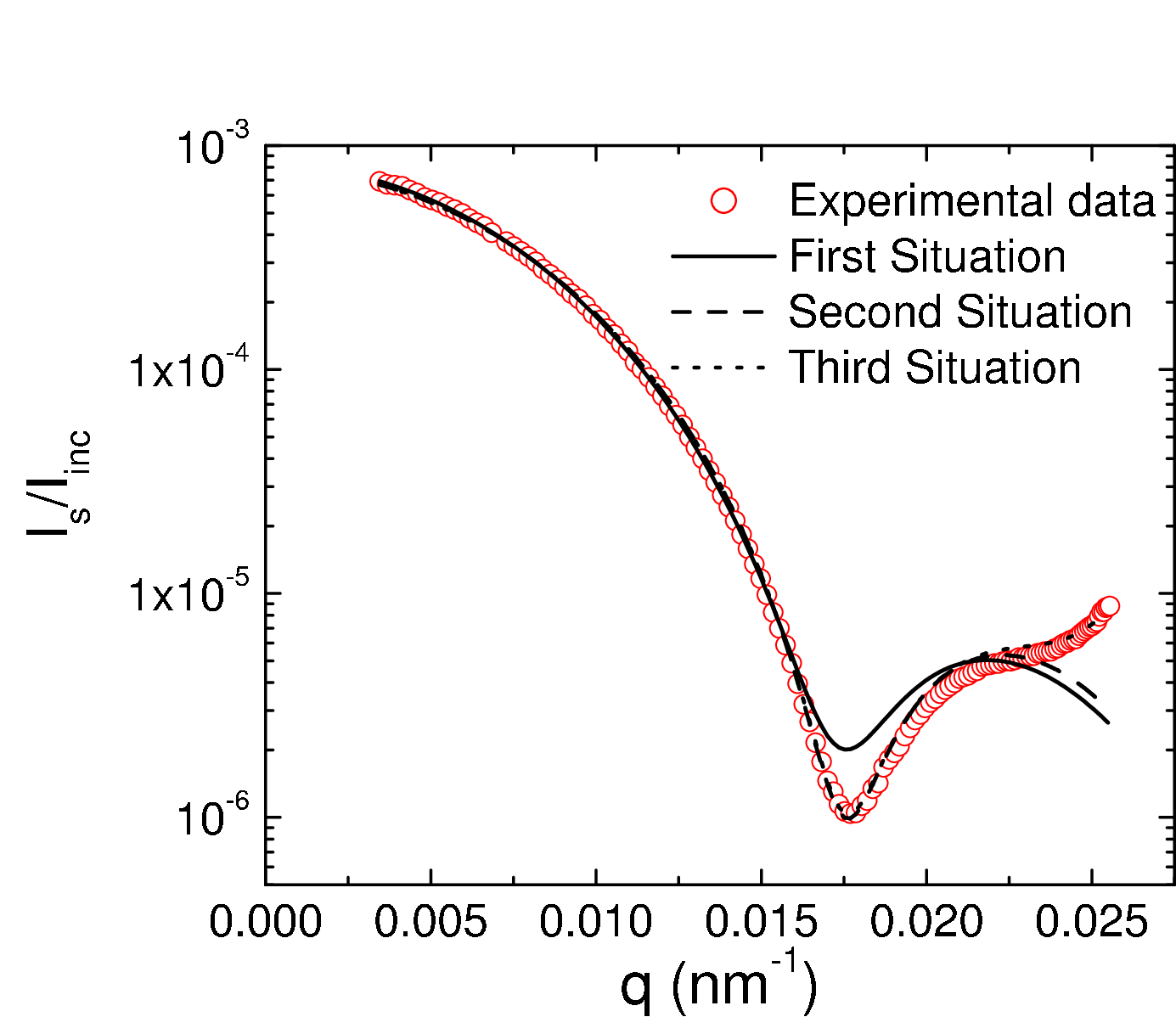}
      \makeatletter\def\@captype{figure}\makeatother
    \caption{The experimental data and expected values for PNIPAM-1 at a temperature
of 302.33 K. The circles show the experimental data, the solid
line shows the expected values calculated using Eq. \ref{mainfit}
in the full particle size distribution range, the dash line
represents the expected values calculated using Eq. \ref{mainfit}
between $\left\langle R_{s}\right\rangle -1.3\sigma $ and
$\left\langle R_{s}\right\rangle +1.3\sigma $ and the dot line
shows the expected values calculated using Eq. \ref{mainre} in the
same range as the second with $b$: $0.014$.}
 \label{figdiffcal}
\end{center}

For the DLS measurements of all the samples investigated, the
expected values calculated based on the particle size distribution
measured using SLS technique or commercial results provided by the
supplier are consistent with the experiment data very well,
respectively. All the results reveal that the static and
hydrodynamic radii are different.

In general, the apparent hydrodynamic radius and polydispersity
index are two important parameters that people try to measure
accurately from DLS measurements. People also believe that the
small values of polydispersity index mean the size distribution of
nanoparticles is narrow or monodisperse and the apparent
hydrodynamic radius is equal to the mean hydrodynamic radius.
Since the apparent hydrodynamic radius and polydispersity index
of particles are obtained through detecting nonexponentility of
the correlation function of temporal fluctuation in the scattered
light at a given scattering angle, so the effects of particle size
distribution and scattering angle on the nonexponentility of the
correlation function of temporal fluctuation in the scattered
light were investigated.

For simplicity, the difference between the static and hydrodynamic
radii of homogeneous spherical particles does not be considered.
The apparent hydrodynamic radius ${R_{app,h}}$ and polydispersity
index $PD.I$ obtained using the Cumulant method at a given
scattering vector $q$ as $\tau\rightarrow0 $ are given by

\begin{equation}
{R_{app,h}}=\frac{\int_{0}^{\infty }R_{s}^{6}P\left( q,R_{s}\right)
G\left( R_{s}\right) dR_{s}}{\int_{0}^{\infty }R_{s}^{5}P\left(
q,R_{s}\right) G\left( R_{s}\right) dR_{s}} \label{Rh}
\end{equation}

\noindent and

\begin{equation}
PD.I=\frac{\int_{0}^{\infty }R_{s}^{4}P\left( q,R_{s}\right)
G\left( R_{s}\right) dR_{s}\int_{0}^{\infty }R_{s}^{6}P\left(
q,R_{s}\right) G\left( R_{s}\right) dR_{s}}{\left(
\int_{0}^{\infty }R_{s}^{5}P\left( q,R_{s}\right) G\left(
R_{s}\right) dR_{s}\right) ^{2}}-1 . \label{Dzindex}
\end{equation}

The values of ${R_{app,h}}$ and $PD.I$ for the mean hydrodynamic
radius 5 nm with different standard deviations at scattering
angles 0$^\mathrm o$, 30$^\mathrm o$, 60$^\mathrm o$, 90$^\mathrm
o$ and 120$^\mathrm o$ are shown in Figs. \ref{figRhPDS}a and
\ref{figRhPDS}b, respectively. For all the calculation, the
wavelength of the incident light in vacuo $\lambda $ is set to
632.8 nm and solvent refractive index $n_{s}$ is 1.332.

\begin{center}
   $\begin{array}{c@{\hspace{0in}}c}
     \includegraphics[width=0.35\textwidth,angle=-90]{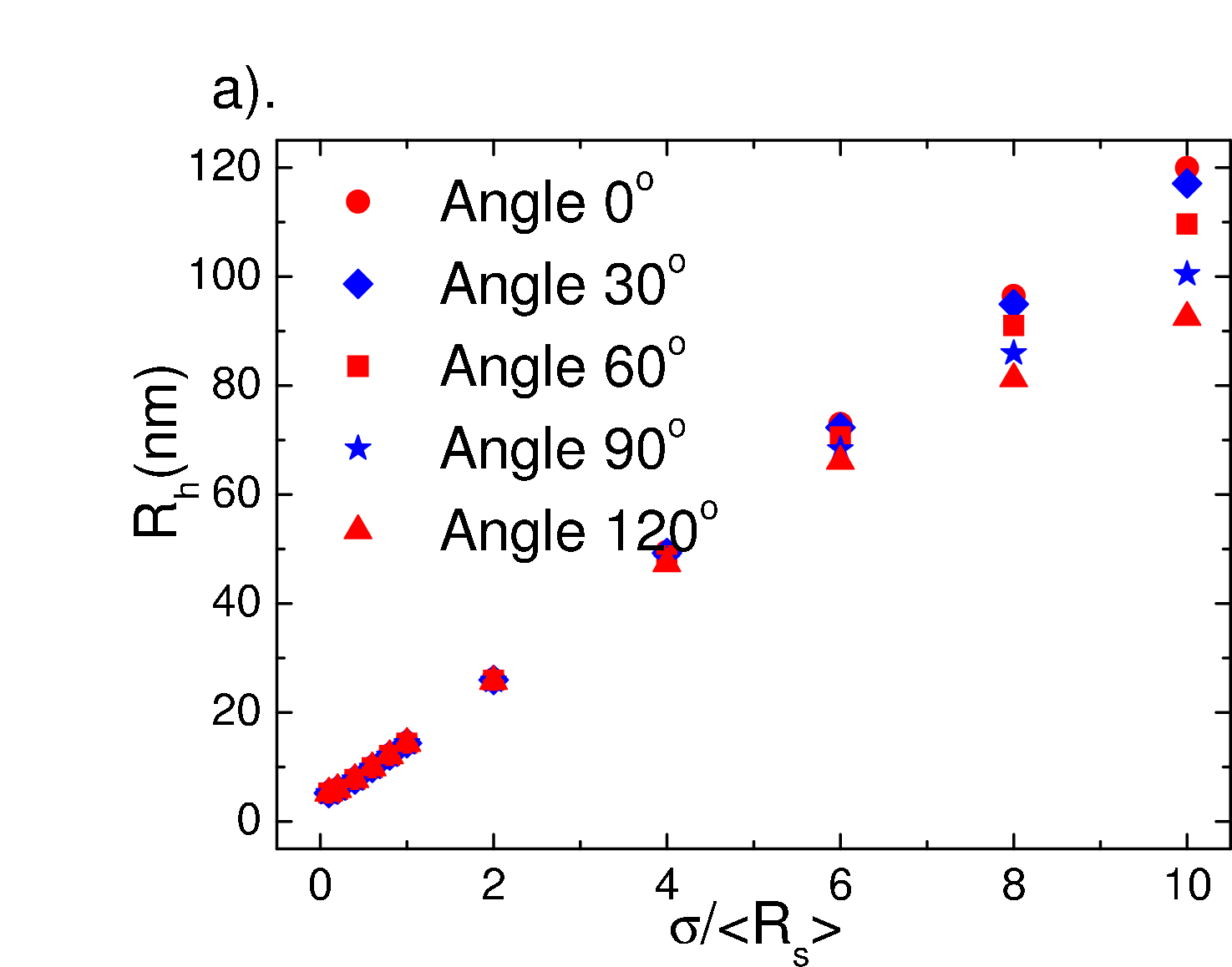} &
     \includegraphics[width=0.35\textwidth,angle=-90]{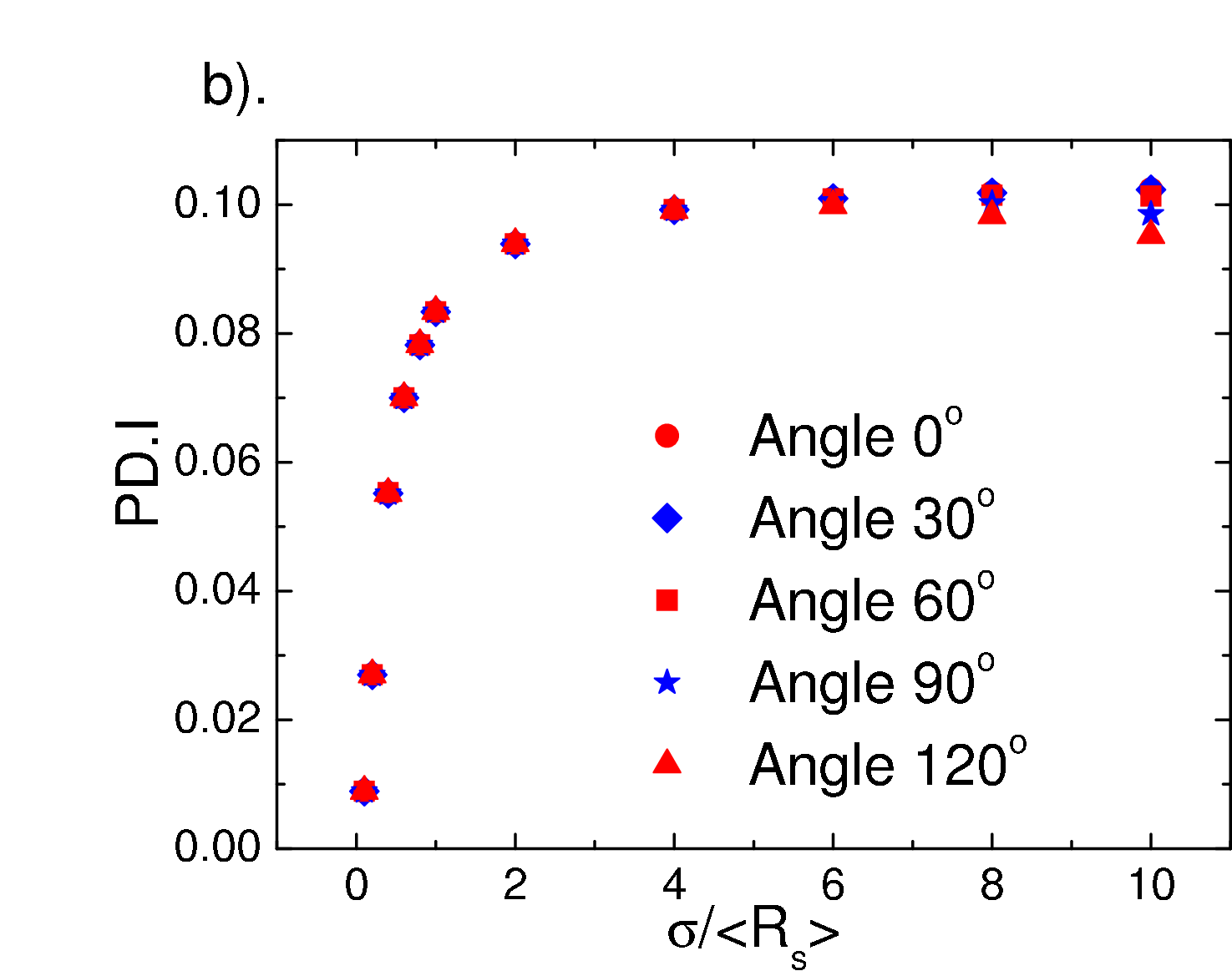} \\ [0.0cm]
   \end{array}$
\end{center}\vspace{-0.5cm}
 \makeatletter\def\@captype{figure}\makeatother
\caption[] {a). The apparent hydrodynamic radius and b).
polydispersity index as a function of relative standard deviation
at scattering angles 0$^\mathrm o$, 30$^\mathrm o$, 60$^\mathrm
o$, 90$^\mathrm o$ and 120$^\mathrm o$ for the nanoparticles with
mean hydrodynamic radius 5 nm.} \label{figRhPDS}

Figure \ref{figRhPDS}a shows that the apparent hydrodynamic radius
is greatly affected by the relative standard deviation and can
have a very large difference with mean hydrodynamic radius. When
the relative standard deviation is less than 1, the effect of
scattering angle can be ignored. At any scattering angle, the
apparent hydrodynamic radius can be measured accurately. Figure
\ref{figRhPDS}b reveals that polydispersity index almost does not
be affected by the scattering angle and approximates a constant
0.1 when the value of relative standard deviation is larger than
3.

The situation for large particles with mean hydrodynamic radius
100 nm also was investigated. ${R_{app,h}}$ and $PD.I$ as a function
of relative standard deviation at scattering angles 0$^\mathrm o$,
30$^\mathrm o$, 60$^\mathrm o$, 90$^\mathrm o$ and 120$^\mathrm o$
are shown in Figs. \ref{figRhPDL}a and \ref{figRhPDL}b,
respectively.

\begin{center}
   $\begin{array}{c@{\hspace{0in}}c}
     \includegraphics[width=0.35\textwidth,angle=-90]{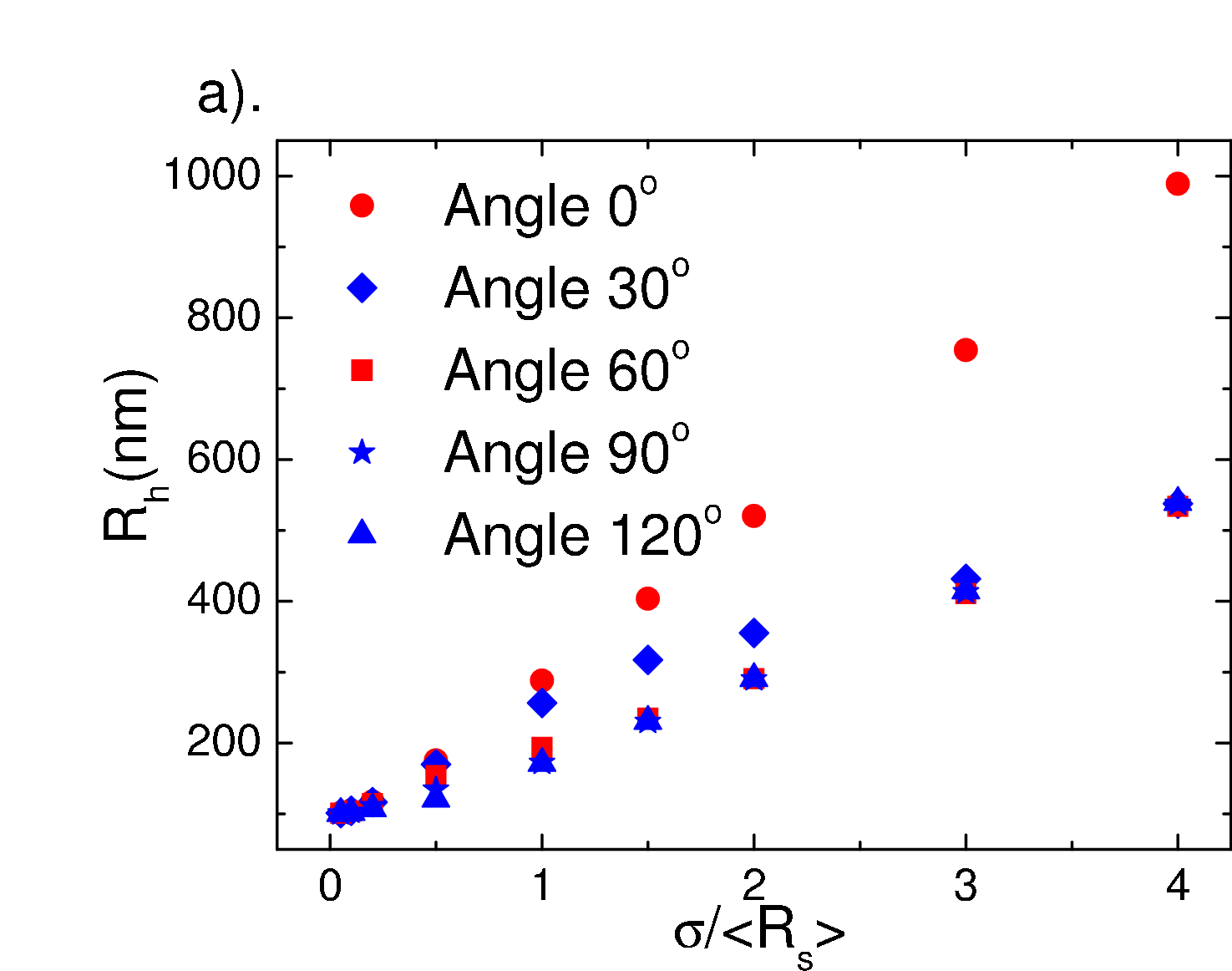} &
     \includegraphics[width=0.35\textwidth,angle=-90]{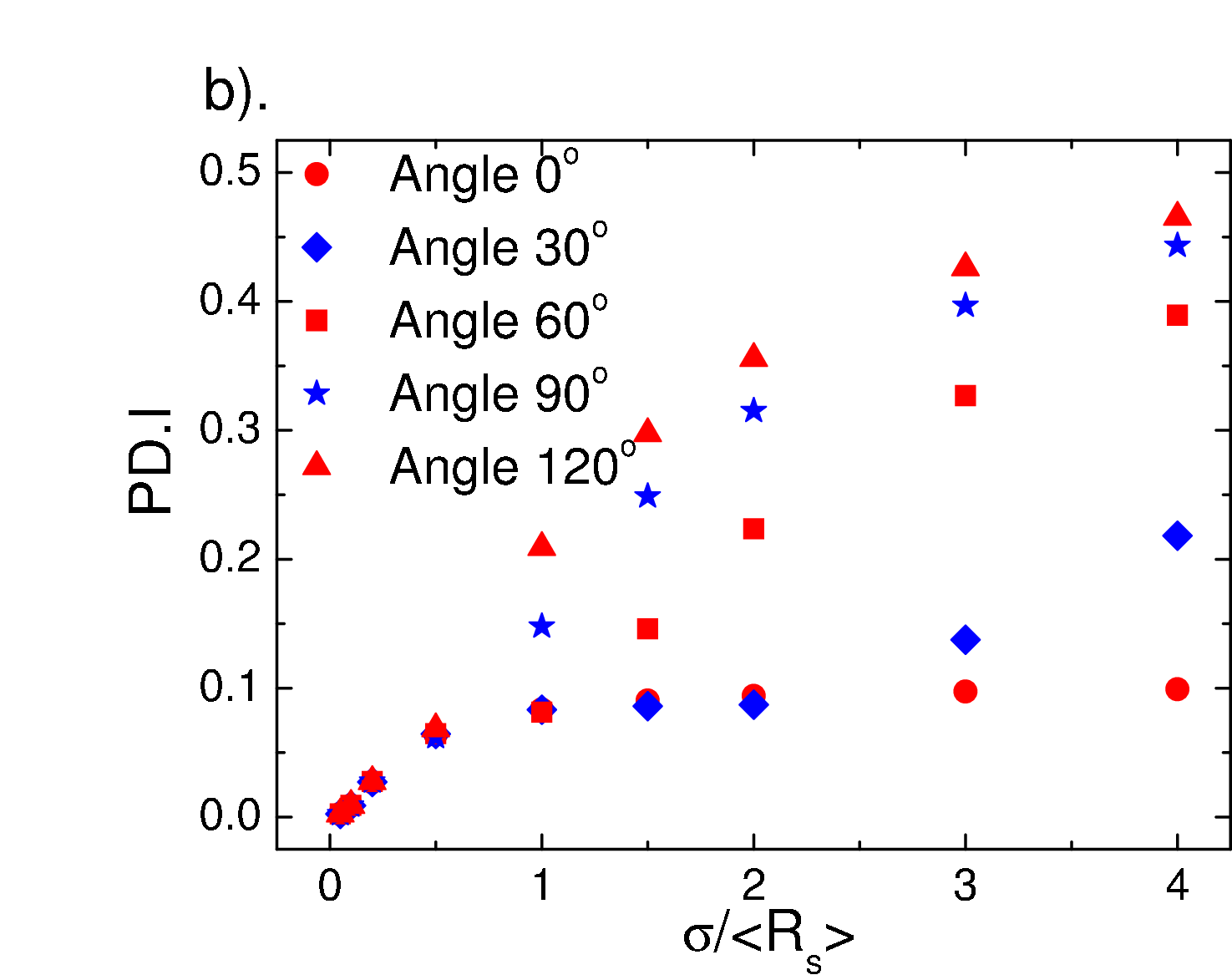} \\ [0.0cm]
   \end{array}$
\end{center}\vspace{-0.5cm}
 \makeatletter\def\@captype{figure}\makeatother
\caption[] {a). The apparent hydrodynamic radius and b).
polydispersity index as a function of relative standard deviation
at scattering angles 0$^\mathrm o$, 30$^\mathrm o$, 60$^\mathrm
o$, 90$^\mathrm o$ and 120$^\mathrm o$ for the nanoparticles with
mean hydrodynamic radius 100 nm.} \label{figRhPDL}

Figure \ref{figRhPDL}a not only shows that the values of apparent
hydrodynamic radius are greatly affected by the relative standard
deviation, but also reveals the complex effects of scattering
angle on ${R_{app,h}}$. Even if the values of apparent hydrodynamic
radius measured at different scattering angles are consistent, it
still does not mean the particle size distribution is narrow or
monodisperse. In order to measure accurately the values of
apparent hydrodynamic radius, the experimental data must be
measured at enough small scattering angles. Figure \ref{figRhPDL}b
shows the same situation as Figure \ref{figRhPDS}b for the
polydispersity index of apparent hydrodynamic radius. As the
relative standard deviation is increased, the polydisperse index
approximates a constant 0.1.

For the polystyrene latex sample, all the results obtained using 
the different techniques are shown in Table \ref{table6}. 
\begin{center}
\begin{tabular}{|c|c|c|c|}
\hline $R_{TEM}$(nm) & $\left\langle R_{s}\right\rangle$(nm) & $R_{app,h}$
(nm) & $R_{app,h}/\left\langle R_{s}\right\rangle$ \\
\hline 33.5 & 33.3$\pm $0.2 & 37.27$\pm $0.09 & 1.119$\pm$0.007 \\
\hline 55 & 56.77$\pm $0.04 & 64.5$\pm $0.6 & 1.14$\pm$0.01 \\
\hline 90 & 92.05$\pm $0.04 & 103.1$\pm $0.4 & 1.120$\pm$0.004 \\
\hline
\end{tabular}
 \makeatletter\def\@captype{table}\makeatother
\caption{The commercial $R_{TEM}$, $\left\langle R_{s}\right\rangle$
apparent hydrodynamic radii.}
\label{table6}
\end{center}
Based on the results above, the sizes obtained using SLS technique are 
consistent with the commercial values obtained using TEM respectively.
The value of the apparent hydrodynamic radius obtained under the same conditions
as the static radius is larger than that of the static radius by about $12\%$.

The values of the root mean square radius of gyration ${\left\langle {R_g}^2\right\rangle}^{1/2}_{cal} $
calculated using the commercial size distribution are consistent with the measured values
of ${\left\langle {R_g}^2\right\rangle}^{1/2}_{Zimm} $ obtained from the Zimm plot analysis. All the results are
shown in Table \ref{table7}.

\begin{center}
\begin{tabular}{|c|c|c|}
\hline Sample & ${\left\langle {R_g}^2\right\rangle}^{1/2}_{cal} $ & ${\left\langle {R_g}^2\right\rangle}^{1/2}_{Zimm} $\\
\hline 33.5(nm) & 26.9  & 26.9$\pm $0.5 \\
\hline 55 (nm)  & 43.24 & 46.8$\pm $0.3  \\
\hline 90 (nm) & 70.1   & 69.0$\pm $2.0 \\
\hline
\end{tabular}
 \makeatletter\def\@captype{table}\makeatother
\caption{Values of ${\left\langle {R_g}^2\right\rangle}^{1/2}_{cal} $ and ${\left\langle {R_g}^2\right\rangle}^{1/2}_{Zimm} $.}
\label{table7}
\end{center}

Based on the discussion above, the size distribution obtained using the SLS is more accurate to represent the
particle information in dispersion and consistent with that measured using the TEM technique. The apparent 
hydrodynamic radius is an optical weight hydrodynamic radius, so it contains much more information of particle
in dispersion comparing with static radius. The ratio of $R_{app,h}/\left\langle R_{s}\right\rangle$ can 
be larger than 2. The results for PNIPAM samples are also shown in Table \ref{table8}. The results also reveal the same situation that the size distribution obtained using the SLS is more accurate to represent the
particle information in dispersion.

\begin{center}
\begin{tabular}{|c|c|c|c|}
\hline Sample (Temperature) & $\rho$  & ${\left\langle {R_g}^2\right\rangle}^{1/2}_{Zimm} /\left\langle R_{s}\right\rangle$ & ${\left\langle {R_g}^2\right\rangle}^{1/2}_{cal}/\left\langle R_{s}\right\rangle $ \\
\hline PNIPAM-5($40^o$C) & 0.73$\pm $0.02 & 0.83$\pm $0.03 & 0.813$\pm$0.003 \\
\hline PNIPAM-2($40^o$C) & 0.69$\pm $0.03 & 0.84$\pm $0.04 & 0.82$\pm$0.01 \\
\hline PNIPAM-1($40^o$C) & 0.69$\pm $0.03 & 0.87$\pm $0.04 & 0.856$\pm$0.009 \\
\hline PNIPAM-0($40^o$C) & 0.66$\pm $0.01 & 0.80$\pm $0.02 & 0.81$\pm$0.01 \\
\hline PNIPAM-0($34^o$C) & 0.54$\pm $0.02 & 1.13$\pm $0.03 & 1.04$\pm$0.03 \\
\hline
\end{tabular}
 \makeatletter\def\@captype{table}\makeatother
\caption{The values of the dimensionless parameters of $\rho$, ${\left\langle {R_g}^2\right\rangle}^{1/2}_{Zimm} /\left\langle R_{s}\right\rangle$ and ${\left\langle {R_g}^2\right\rangle}^{1/2}_{cal}/\left\langle R_{s}\right\rangle $.}
\label{table8}
\end{center}

\section{CONCLUSION}

Using the Eq. \ref{mainfit}, the accurate size distribution of static radii $R_{s}$
can be measured using the SLS technique. Since the size distributions of polystyrene latex samples
obtained using the TEM and SLS techniques are consistent and the values of the root mean square radius of gyration ${\left\langle {R_g}^2\right\rangle}^{1/2}_{cal} $
calculated using the commercial size distribution or static size distributions are consistent with the measured values
of ${\left\langle {R_g}^2\right\rangle}^{1/2}_{Zimm} $ obtained using the Zimm plot analysis respectively,
the size distribution obtained using the SLS is more accurate to represent the
particle information in dispersion. The static and hydrodynamic radii describe the distinct characteristics of particles
in dispersion, so they are different physical quantities and can have very large difference. With the SLS and DLS techniques together, the much more information about the particles in dispersion can be explored further. Using the SLS technique to study different shape particles in dispersion, the results of research maybe can improve our understanding for the scattering light from the particles in dispersion totally. The apparent hydrodynamic radius and polydispersity index cannot give an accurate description for size distribution.

\end{document}